\documentclass[12pt]{article}
\usepackage{eurosym}
\usepackage{graphicx}
\usepackage{amsthm}
\usepackage{amsmath}
\usepackage{amssymb}
\usepackage{appendix}
\usepackage{color}
\usepackage{sgame}
\usepackage{verbatim}
\usepackage{caption}
\usepackage{subcaption}
\usepackage{mathtools}
\usepackage{dsfont}
\usepackage{float}
\usepackage[margin=1in]{geometry}
\usepackage{natbib}
\usepackage[utf8]{inputenc}
\usepackage[english]{babel}
\usepackage{lscape}
\usepackage[colorlinks=true,allcolors=blue]{hyperref}
\usepackage{booktabs}
\usepackage{longtable}
\usepackage{threeparttable}
\usepackage{dcolumn}
\usepackage{tikz}
\usepackage{xcolor}
\usepackage{istgame}
\usepackage{multirow}
\usepackage{amsfonts}
\usepackage{adjustbox}
\usepackage{istgame}
\usepackage{bbm}
\usepackage{soul}

\usetikzlibrary{calc}
\usetikzlibrary{calc}
\newtheorem{corollary}{Corollary}

\newtheorem{definition}{Definition}

\newtheorem{proposition}{Proposition}

\newtheorem{remark}{Remark}

\DeclareMathOperator*{\argmax}{argmax}
\begin{document}

\title{Cursed Job Market Signaling\thanks{Support from the 
Quantitative Collaboration at the University of Virginia is gratefully 
acknowledged. We thank Meng-Jhang Fong, Menglong Guan,
Charles Holt, Shengwu Li, John Lightle, Alexander MacKay, Thomas R. Palfrey, 
Régis Renault, Emanuel Vespa, Pin-Chao Wang, Stephanie W. Wang and 
audiences at University of Virginia, National Taiwan University and the 2025 ESA North American Conference
for helpful comments. We are also grateful to 
Dorothea Kübler, Wieland Müller and Hans-Theo Normann for sharing their experimental data with us.}}
\author{Po-Hsuan Lin\thanks{%
Corresponding Author: Department of Economics, University 
of Virginia, Charlottesville, VA 22904. Email: 
plin@virginia.edu} \and 
Yen Ling Tan\thanks{%
Department of Economics, University of Virginia, Charlottesville, VA 22904. Email: 
yt2dh@virginia.edu}}
\date{\today}
\maketitle

\begin{abstract}

We study how \emph{cursedness}, the tendency to neglect how other people's 
strategies depend on their private information, affects information 
transmission in Spence's job market signaling game. We characterize 
the Cursed Sequential Equilibrium and show that as players 
become more cursed, 
the worker obtains less education---a costly signal 
that does not enhance productivity---suggesting that cursedness improves 
the efficiency of information transmission. However, this efficiency 
improvement depends on the richness of the message space. 
Revisiting the job market signaling experiment by Kübler, Müller, and Normann
(2008), we find evidence consistent with our theory. Finally, we show that 
cursedness has implications for wage compression in labor markets.

\end{abstract}

\renewcommand{\baselinestretch}{1.2}

\thispagestyle{empty}

\bigskip



\bigskip

Word Count: 6363

\newpage\setcounter{page}{1} \renewcommand{\baselinestretch}{1.2}

\section{Introduction}

Many economic and political settings with incomplete information can be modeled as signaling games, including education \citep{spence1973job}, insurance \citep{rothschild1976insurance}, limit pricing \citep{milgrom1982limit}, and leadership \citep{hermalin1998toward}. In these settings, a privately informed sender takes a costly action that an otherwise uninformed receiver uses to infer information about the sender that would not be directly observable, such as their ability or the quality of their product. A natural question is when such actions can credibly convey the sender's private information to the receiver.

In standard models, achieving full information transmission requires the receiver to understand how the sender's message depends on their private type, and the sender to anticipate that the receiver understands this dependence. This requirement is demanding: a large body of experimental evidence shows that people often fail to account for how an opponent's strategy depends on information observed only by the opponent. One of the clearest illustrations comes from common-value auctions, where bidders tend to overbid because they fail to recognize that winning the auction is itself informative about how little their rivals valued the object---a pattern known as the \emph{winner's curse} \citep{capen1971competitive,kagel1986winner,hendricks2003empirical}. We use \emph{cursedness} as a general term for this underlying cognitive bias, which has also been documented in a range of strategic settings, including bilateral bargaining games \citep{samuelson1985negotiation, holt1994loser, carrillo2011no}, zero-sum betting games with asymmetric information \citep{rogers2009heterogeneous}, voting \citep{guarnaschelli2000experimental, esponda2014hypothetical}, adverse selection settings \citep{charness2009,martinez2019failures}, and cheap talk games \citep{lim2024controlled}. In the signaling game environment central to this paper, \cite{brandts1993adjustment} find that even after many repetitions, senders' and receivers' choices still deviate from the predictions of the standard equilibrium model, in the direction consistent with cursedness.

How does cursedness affect information transmission in signaling games?
To answer this question in a parsimonious way, we adopt the Cursed Sequential 
Equilibrium (CSE) proposed by \cite{fong2023cursed} as our solution concept.
CSE extends the classic Cursed Equilibrium (CE) by \cite{eyster2005}
from static Bayesian games to multi-stage games of incomplete information.
Like CE, CSE assumes that players partially 
neglect the correlation between others' behavioral strategies
and their private information at each stage of the game.\footnote{
CSE is not the only solution concept that extends CE from static Bayesian games
to dynamic games. \cite{cohen2022sequential} 
extend CE to general extensive games with perfect recall by considering a different source of cursedness, called the Sequential Cursed Equilibrium (SCE); for a detailed comparison, see  \cite{fong2023notecursedsequentialequilibrium}.} 
Specifically, the model is parameterized by a single parameter, $\chi\in[0,1]$. 
A $\chi$-cursed player incorrectly believes that, with probability $\chi$, 
the opponents adopt the average behavioral strategies regardless of their private types, 
while with probability $1-\chi$, the opponents adopt the true type-dependent behavioral strategies.
Therefore, a higher $\chi$ means players are \emph{more cursed}, 
as their perception of other players' behavioral strategies becomes increasingly distorted.

In Section \ref{sec:cse_label}, we formally define CSE for a general class of signaling games. The key differences between CSE and CE are that CSE is an equilibrium concept in behavioral strategies 
(rather than normal-form mixed strategies) and 
requires sequential rationality.
Furthermore, for off-path events, CSE imposes a $\chi$-consistency requirement. This requirement is analogous to the consistency condition for sequential equilibrium in \cite{krepswilson1982}, where a consistent assessment 
is approachable by a sequence of assessments with totally mixed behavioral strategies. The $\chi$-consistency requirement 
plays a crucial role in our analysis of signaling games.

As a first step in studying the effect of cursedness in signaling games,
we focus on one of the simplest applications: the job market signaling
model by \cite{spence1973job}. In this model, a worker (the sender) privately 
knows whether their productivity is high or low and can obtain costly 
education---which does not affect their productivity---as a signal to a 
firm (the receiver). Since education does not enhance 
productivity, transmitting the same amount of information 
with less education constitutes an efficiency improvement.

In Section \ref{sec:spence_signaling}, we characterize the set of $\chi$-CSE, 
showing that as long as players are not fully cursed (i.e., $\chi < 1$), a 
separating $\chi$-CSE always exists (Proposition \ref{prop:sep_cse_ic}), 
regardless of the degree of cursedness. However, when players are fully 
cursed ($\chi = 1$), the sender believes that the receiver completely ignores the informational content
of the costly signal, leading both sender types to choose
zero education (Proposition \ref{prop:pool_cse_ic}).
This result suggests that the effect of cursedness on information
transmission is discontinuous at $\chi = 1$.
As long as the receiver believes the sender’s signal is 
informative, even if only slightly, and the sender
shares this belief, a separating $\chi$-CSE can be sustained.

There are multiple separating and pooling $\chi$-CSE for $\chi < 1$. 
In Corollaries \ref{coro:monotone_separating} and \ref{coro:monotone_pooling},
we show that the sets of education levels in 
separating and pooling $\chi$-CSE decrease 
with $\chi$, 
implying that when players are more cursed, the worker attains less 
education, i.e., send less costly signals, in equilibrium.\footnote{This comparative static is in the sense of the weak set order 
introduced by \cite{che2019weak}. See Section
\ref{sec:spence_signaling} for a detailed description of the weak set order.}
This result shows that in the canonical job market signaling model, 
cursedness \emph{improves} the efficiency of information transmission,
as the same amount of information is transmitted with less 
socially wasteful signaling.

The rationale behind this qualitative result stems from 
the $\chi$-consistency requirement of $\chi$-CSE. When
players are $\chi$-cursed, an implication of $\chi$-consistency is
that the receiver's posterior belief about any type is
bounded below by $\chi$ times the prior probability of that type, 
regardless of whether the signal is on-path or off-path (Proposition \ref{prop:chi_dampen_update}).
Hence, as $\chi$ increases, the sender \emph{over-estimates} the expected
payoff of deviating from the equilibrium 
signal, making the incentive compatibility (IC) condition more stringent and leading to a
lower equilibrium education level.

Although $\chi$-consistency imposes restrictions on off-path beliefs,
it is not strong enough to pin down a unique prediction in signaling games.
To further refine the set of multiple $\chi$-CSE, we propose the
Cursed Intuitive Criterion in Section \ref{sec:cursed_intu_criter}.
Similar to the Intuitive Criterion of \cite{cho1987signaling}, our criterion
requires that, upon observing an off-path message,
players assign the lowest probabilities---consistent with
$\chi$-consistency---to the types of senders who have no incentive to send the 
off-path message, as it is equilibrium-dominated. 
Since the cursed intuitive criterion is defined to be compatible with
$\chi$-consistency, equilibrium selection is, in principle, highly sensitive to the degree of cursedness.  
However, by applying our cursed intuitive
criterion to refine the $\chi$-CSE of the job market signaling game, 
Proposition \ref{prop:job_market_selection} shows that, for any $\chi < 1$, it always selects the most efficient (in the sense 
of Pareto-dominance) separating
$\chi$-CSE,\footnote{When $\chi = 1$,
the cursed intuitive criterion selects the unique pooling CSE where
both types of the worker choose zero education. This equilibrium is 
the most efficient pooling equilibrium.}
which corresponds to the Riley outcome \citep{riley1975competitive, riley1979informational}. 
This result suggests that the Riley outcome is robust to cursedness.

To examine whether our cursed intuitive criterion better 
explains the experimental data than the standard intuitive criterion, 
we revisit the job market signaling experiment of \cite{kubler2008job} 
in Section \ref{sec:experimental_evidence}. In this experiment, there are 
two productivity types, and education is restricted to a binary choice 
rather than a continuum of choices. While the prediction of the intuitive criterion 
is invariant to this restriction, the cursed intuitive criterion makes 
a distinct prediction. It suggests that when players are sufficiently 
cursed, the restricted message space may prevent any information transmission, 
resulting in a pooling equilibrium where none of the types invest in education. 
In the experiment, even with many repetitions, subjects do not fully converge
to the prediction of the intuitive criterion, instead deviating in the direction 
predicted by the cursed intuitive criterion. This finding provides
supportive evidence for our theory.

Finally, in Section \ref{sec:conclusion}, we apply our theory to the job market signaling game with a continuum of types to shed light on \emph{wage compression}---the 
well-documented tendency for the wage gap between workers to be 
narrower than the underlying gap in their productivity. We show that cursedness alone, independent of factors such as fairness norms or institutional wage rigidities, compresses wages, offering a novel channel for this empirical regularity in labor markets.

\section{Cursed Sequential Equilibrium in Signaling Games}
\label{sec:cse_label}

For simplicity, this section defines the cursed sequential 
equilibrium (CSE) within a general class of signaling games. 
CSE is more broadly defined for multi-stage games of incomplete information. 
For a detailed description, see \cite{fong2023cursed}.

A signaling game consists of two players: a sender (player 1) and a receiver (player 2). The sender has a private type $\theta \in \Theta$, drawn from a common prior distribution
where $F(\theta)>0$ for all $\theta\in \Theta$. After observing their type, the sender chooses a message $m \in M$. Upon receiving the message, the receiver then selects an action $a \in A$.\footnote{For simplicity, 
we assume that the action space is independent of the messages.}
The payoff functions for the sender and the receiver are represented by von Neumann-Morgenstern utility functions, $u_1(\theta, m, a)$ and $u_2(\theta, m, a)$, respectively.\footnote{Since $u_1(\cdot)$ and $u_2(\cdot)$ are von Neumann-Morgenstern utility functions, they can be extended to the strategy spaces associated with $\Delta(M)$ and $\Delta(A)$ by taking expected values.}
A behavioral strategy for the sender, $(\sigma_1(\cdot | \theta))_{\theta \in \Theta}$, assigns a probability distribution over $M$ for each type. Similarly, a behavioral strategy for the receiver, $(\sigma_2(\cdot | m))_{m \in M}$, assigns a probability distribution over $A$ for each message.

The cursed sequential equilibrium of a signaling game is defined as an \emph{assessment} consisting of a behavioral strategy profile $\sigma = (\sigma_1, \sigma_2)$ and a belief system $\mu = (\mu(\cdot | m))_{m \in M}$, which assigns a probability distribution over $\Theta$ upon receiving each message $m$. In CSE, a cursed player fails to recognize how other players' actions depend on their private types. Specifically, 
in signaling games, since the sender is the only player with private information, 
the receiver is the only one who has to infer their opponent's type.

To formally model cursedness in signaling games, for any assessment $(\mu, \sigma)$, we define the \emph{average behavioral strategy of the sender} as
$$\bar{\sigma}_1(m) = \sum_{\theta\in\Theta} F(\theta)\sigma_1(m | \theta).$$
In CSE, the receiver is assumed to have incorrect 
perceptions of the sender's behavioral strategy. Instead of believing 
that the sender is using $\sigma_1$, a $\chi$-cursed receiver perceives 
the sender as following a $\chi$-weighted average of the 
true behavioral strategy and the average behavioral strategy:
$$\sigma_1^\chi(m|\theta) = \chi \bar{\sigma}_1(m) + (1-\chi)\sigma_1(m|\theta).$$

The receiver's belief about the sender's private type is updated in a
$\chi$-CSE using Bayes’ rule whenever possible, under the assumption 
that the sender follows the $\chi$-cursed
behavioral strategy.
This updating rule, called the \emph{$\chi$-cursed Bayes' rule}, is defined as follows.

\begin{definition}\label{definition:cursed_bayes_rule}
An assessment of a signaling game $(\mu, \sigma)$ satisfies $\chi $-cursed Bayes'
rule if the following is applied to update the receiver's belief whenever
$\sum_{\theta\in\Theta} F(\theta)\sigma_1^\chi(m|\theta)>0$:
$$\mu(\theta | m) = \frac{F(\theta)\sigma_1^\chi(m|\theta)}{\sum_{\theta'\in\Theta} F(\theta')\sigma_1^\chi(m|\theta')}. $$
\end{definition}

Additionally, CSE imposes a consistency requirement on how 
$\chi$-cursed beliefs are updated off the equilibrium path, i.e., when
$\sum_{\theta\in\Theta} F(\theta)\sigma_1^\chi(m|\theta)=0$. Let $\Sigma^0$ 
denote the set of totally mixed behavioral strategy profiles, and let $\Psi ^{\chi }$
be the set of assessments $(\mu, \sigma)$ such that $\sigma \in \Sigma^0$ and 
$\mu$ is derived from $\sigma$ using $\chi$-cursed Bayes' rule.
An assessment satisfies $\chi$-consistency if it belongs to the
closure of $\Psi ^{\chi }$, denoted as $\mbox{cl}(\Psi ^{\chi })$.

\begin{definition}\label{definition:chi_consistency}
An assessment of a signaling game $(\mu, \sigma)$ satisfies $\chi $-consistency
if there is a sequence of assessments $\{(\mu^{k},\sigma^{k})\}\subseteq
\Psi ^{\chi }$ such that $\lim_{k\rightarrow \infty }(\mu^{k}, \sigma^{k})=
(\mu ,\sigma )$.
\end{definition}

Propositions \ref{prop:curse_bayes_rule} and \ref{prop:chi_dampen_update} 
state two important belief-updating properties of $\chi$-CSE. The proofs 
of these properties can be found in \cite{fong2023cursed}. In summary, 
the $\chi$-cursed Bayes' rule uniquely determines the receiver's 
posterior belief for all on-path messages. One implication 
of this is that the receiver's posterior belief for any sender type $\theta$, 
conditional on any message $m$, is bounded below by $\chi F(\theta)$. 
Moreover, $\chi$-consistency requires that a $\chi$-consistent assessment 
to be approachable by a sequence of assessments in $\Psi^{\chi}$
with totally mixed behavioral strategies. As a result, for
any $\chi$-consistent assessment, the receiver's posterior belief 
for any sender type $\theta$, conditional on any message $m$, 
is also bounded below by $\chi F(\theta)$, \emph{regardless of
whether $m$ is on-path or off-path.}

\begin{proposition}\label{prop:curse_bayes_rule}
For any assessment of a signaling game $(\mu, \sigma)\in \Psi^\chi$,
$m\in M$ and $\theta\in\Theta$,
$$\mu(\theta | m) = \chi F(\theta) + (1-\chi)\left[\frac{F(\theta)\sigma_1(m|\theta)}{\sum_{\theta'\in\Theta}F(\theta')\sigma_1(m|\theta')} \right].$$
\end{proposition}

\begin{proposition}\label{prop:chi_dampen_update}
For any assessment of a signaling game $(\mu, \sigma)\in \mbox{\emph{cl}}(\Psi ^{\chi })$, 
$m\in M$ and $\theta\in\Theta$,
$$\mu(\theta | m) \geq \chi F(\theta).$$    
\end{proposition}

Lastly, $\chi$-CSE of a signaling game is defined as below.

\begin{definition}\label{definition:chi_cse}
An assessment of a signaling game $(\mu, \sigma)$ is a $\chi$-cursed sequential 
equilibrium if it satisfies $\chi$-consistency and
\begin{itemize}
    \item[(i)] for each $\theta\in \Theta$ and $m^*\in M$ 
    such that $\sigma_1(m^*|\theta)>0$, 
    $$m^*\in \argmax_{m\in M} \sum_{a\in A} u_1(\theta, m, a)\cdot\sigma_2(a|m);$$
    \item[(ii)] for each $m\in M$ and $a^*\in A$ such that $\sigma_2(a^*|m)>0$,
    $$a^*\in \argmax_{a\in A} \sum_{\theta\in \Theta}
    u_2(\theta, m, a)\cdot \mu(\theta| m).$$
\end{itemize}
\end{definition}

\section{Spence's Job Market Signaling Game}
\label{sec:spence_signaling}

This paper characterizes the cursed sequential equilibrium of an important signaling game: 
the job market signaling game introduced by \cite{spence1973job}. To illustrate the effect of cursedness, we first consider the simplest version: one worker (the sender) with two types and one firm (the receiver)

The worker has two private types, $\Theta=\{\theta_L,\theta_H\}$, with $\theta_H>\theta_L$, where the types can 
be interpreted as the worker's ability. The prior probability of the worker being type $\theta_H$ is given by 
$F(\theta_H) = p>0$.
After learning their type, the worker chooses an education level $e\in M = \mathbb{R}_+$ 
at a cost $c(e|\theta)$. We assume the cost function 
is twice differentiable and has the following properties: 
(i) $c(0|\theta)=0$ for all $\theta$, (ii) the cost is strictly increasing 
and convex in education, i.e., $c'(e|\theta)>0$ and $c''(e|\theta)>0$ for all $e$ and $\theta$, and (iii) the 
high-type worker has a smaller marginal cost, i.e., $c'(e|\theta_H) < c'(e|\theta_L)$ for all $e$.
Following \cite{spence1973job}, we assume the worker's education level has no effect on 
their productivity.

The firm does not observe the worker's type but can observe the worker's education level. 
Upon observing the worker's education level, the firm offers a wage 
$w\in A = \mathbb{R}_+$ to the worker. The firm's payoff function is given by 
$u_2(\theta, e, w) = -(w - \theta)^2$, meaning the firm's objective is to minimize the 
quadratic difference between the wage and the worker's productivity. Therefore, the firm 
offers a wage equal to the expected productivity in equilibrium.\footnote{The 
quadratic loss function captures the outcome of Bertrand competition among multiple firms.
Alternatively, we could assume that several firms compete and make simultaneous wage offers to a worker, with the
firms' payoff functions given by $u_2(\theta,e,w) = \theta-w$.}
Lastly, the worker's payoff function is $u_1(\theta, e, w) = w - c(e|\theta)$.

There are multiple sequential equilibria in the canonical Spence job market signaling game, including both separating and pooling equilibria. In any separating equilibrium, the firm can perfectly infer the worker's type from the observed education level. As a result, in equilibrium, the type $\theta_L$ worker acquires no education, while the type $\theta_H$ worker chooses a sufficiently high level of education to deter the type $\theta_L$ worker from mimicking their strategy. On the other hand, in any pooling equilibrium, where both worker types choose the same education level, the firm offers the average wage on path because it cannot distinguish between the two types. Moreover, upon observing any off-path education level, the firm pessimistically believes that the worker is of type $\theta_L$ and offers the low-productivity wage, thereby supporting the pooling equilibrium.

This equilibration relies on the firm correctly perceiving how a worker's education choice depends on their private type. When players fail to recognize this dependence, one might expect that no information is transmitted. Surprisingly, however, this is not the case. In Proposition \ref{prop:sep_cse_ic}, 
we show that, as long as players are not fully cursed,
there exists a separating $\chi$-CSE. This implies that information transmission is
always possible as long as $\chi < 1$. Furthermore, Proposition \ref{prop:pool_cse_ic}
characterizes the set of pooling $\chi$-CSE, showing that cursedness \emph{refines} pooling 
equilibria by imposing additional restrictions on off-path beliefs.

\begin{proposition}\label{prop:sep_cse_ic}
In any separating $\chi$-CSE of the job market signaling game where 
$e_L \neq e_H$, 
\begin{itemize}
    \item[1.] the type $\theta_L$ worker will choose $e_L = 0$, 
    and 
    \item[2.] the type $\theta_H$ worker will choose $e_H>0$ such that $c(e_H|\theta_L)\geq(1-\chi)(\theta_H-\theta_L)\geq c(e_H|\theta_H)$.
\end{itemize}
\end{proposition}

Proposition \ref{prop:sep_cse_ic} characterizes the set 
of separating $\chi$-CSE for any $\chi\in[0,1]$. When 
$\chi=0$, the IC conditions simplify to
$c(e_H|\theta_L)\geq \theta_H-\theta_L
\geq c(e_H|\theta_H)$, which corresponds to the
result of the standard model. 
At the other extreme, when $\chi=1$, 
the IC conditions reduce to $c(e_H|\theta_H)=0 
\iff e_H=0$ by strict monotonicity of the cost function. 
In other words, when players are fully cursed, 
there is no separating equilibrium. 
For intermediate values of $\chi \in (0,1)$, 
there always exists a continuum of $e_H$ that support a separating $\chi$-CSE, suggesting
that no matter how cursed the players are,
information transmission is always achievable. Additionally, 
we can characterize how the set of separating $\chi$-CSE 
changes monotonically with $\chi$ using the 
weak set order introduced by \cite{che2019weak}.

\begin{definition}\label{def:weak_set_order}
For any two sets $S$ and $S'$, $S'$ weak set dominates 
$S$, denoted by $S' \geq_{ws} S$,
if for any $x\in S$, there 
is $x'\in S'$ such that $x' \geq x$ and for any $x'\in S'$,
there is $x\in S$ such that $x' \geq x$.
\end{definition}

We use $\mathcal{S}^\chi$
to denote the set of education levels 
chosen by the type $\theta_H$ worker in a separating 
$\chi$-CSE. That is, 
$\mathcal{S}^\chi \equiv \left\{e_H :
c(e_H|\theta_L)\geq(1-\chi)(\theta_H-\theta_L)\geq 
c(e_H|\theta_H) \right\}$. In Corollary~\ref{coro:monotone_separating}, we show that 
$\mathcal{S}^\chi$ decreases in $\chi$
under the weak set order, implying that as players become more 
cursed, the high type worker attains a 
lower education level in 
a separating $\chi$-CSE.\footnote{Here we emphasize that this comparative static holds in the sense of the weak set order, since multiple equilibria exist. The result under the unique selection of $\chi$-CSE is established later in Section \ref{sec:cursed_intu_criter}.}
The intuition is that as players become more cursed, the firm's beliefs become ``stickier'' to their prior and adjust less in response to the worker's education choice, making signaling less effective. Consequently, the high type worker chooses a less costly signal in a separating equilibrium.

\begin{corollary}\label{coro:monotone_separating}
For any $\chi, \chi' \in [0,1)$, if $\chi \geq \chi'$, 
then $\mathcal{S}^{\chi'} \geq_{ws} \mathcal{S}^\chi$.
\end{corollary}

Although the efficiency of information transmission
increases as players become more cursed, the high type worker chooses a lower 
level of education in equilibrium. This creates a tradeoff: while cursedness can theoretically 
improve the efficiency of information transmission, separating $\chi$-CSE may be more 
difficult to sustain in practice. In particular, in noisy or inattentive environments, 
the smaller separation in education levels makes the two types harder for the firm to distinguish.
Hence, we conjecture that in such environments, players are more likely to sustain pooling equilibria, which we characterize next.

Now, consider any pooling $\chi$-CSE, where $e_L = e_H = e^*$. Since both types choose the same education level, cursedness does not affect on-path beliefs: the firm's posterior equals its prior. In contrast, $\chi$-consistency imposes additional restrictions on off-path beliefs, which shape the set of pooling $\chi$-CSE as $\chi$ varies, as shown in Proposition \ref{prop:pool_cse_ic}.

\begin{proposition}\label{prop:pool_cse_ic}
In any pooling $\chi$-CSE of the job market signaling game where $e_L = e_H = e^*$, 
$$c(e^*| \theta_L) \leq (1-\chi)p(\theta_H - \theta_L).$$
\end{proposition}

From Proposition \ref{prop:pool_cse_ic}, we find that as $\chi$ increases, the equilibrium restriction becomes more stringent, causing the set of education levels that can be supported as pooling $\chi$-CSE to shrink. In other words, cursedness \emph{refines} the set of pooling equilibria.
This result sharply contrasts with the prediction of cursed 
equilibrium in \cite{eyster2005}, who show that any pooling Bayesian equilibrium 
is a $\chi$-CE for any $\chi \in [0,1]$. In other words,
the set of pooling $\chi$-CE coincides with the set of pooling Bayesian equilibria. This 
sharp difference between CSE and CE is driven by $\chi$-consistency, which restricts the set of 
admissible off-path beliefs and thereby restricts the set of 
pooling equilibria.\footnote{In contrast, CE and CSE coincide for separating equilibria: the cursed firm holds the same distorted perception of the worker's strategy under both solution concepts, and $\chi$-consistency imposes no additional restriction in this case.}

To formally characterize how the set of pooling $\chi$-CSE varies
with $\chi$, we define $\mathcal{P}^\chi$ as the set of 
education levels that can be supported in a pooling $\chi$-CSE. That is, 
$\mathcal{P}^\chi\equiv\{e: c(e|\theta_L)\leq (1-\chi)p(\theta_H - \theta_L) \}$. 
Corollary \ref{coro:monotone_pooling} proves that  
the set of pooling $\chi$-CSE decreases in $\chi$ under
the weak set order. Furthermore, when $\chi=1$,
the unique CSE is the pooling equilibrium where both types choose $e=0$,
which is the most efficient pooling equilibrium.

\begin{corollary}\label{coro:monotone_pooling}
For any $\chi, \chi'\in [0,1]$, if $\chi\geq \chi'$, then $\mathcal{P}^{\chi'}
\geq_{ws} \mathcal{P}^\chi$. Moreover,  the unique fully cursed sequential equilibrium 
$(\chi=1)$ is the pooling equilibrium where $e_L = e_H = 0$.
\end{corollary}

\begin{figure}[H]
\centering
\includegraphics[width=0.9\textwidth]{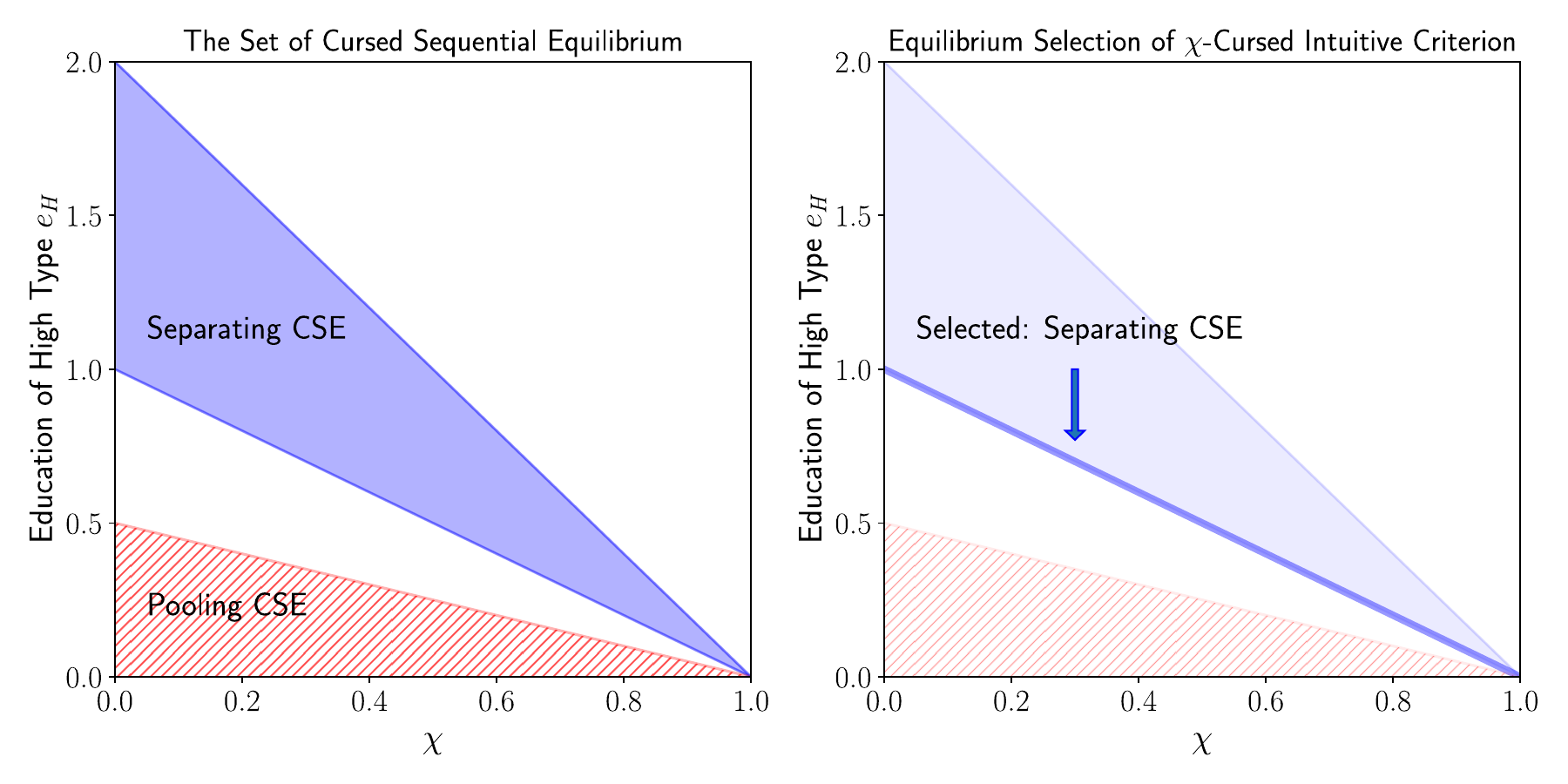}
\caption{(Left) The set of $\chi$-CSE for $c(e|\theta)=e/\theta$, 
$\theta_H=2$, $\theta_L=1$, and $p=0.5$. (Right) Equilibrium selection 
under the $\chi$-cursed intuitive criterion.}
\label{cse}
\end{figure}

To summarize, we present a numerical example where
$c(e|\theta) = e/\theta$, $\theta_H = 2$, $\theta_L = 1$, and $p = 0.5$.
The left panel of Figure \ref{cse} illustrates the set of 
separating and pooling $\chi$-CSE, with
the horizontal axis representing the degree of cursedness $\chi$ and the
vertical axis showing the
education level of the type $\theta_H$ worker in equilibrium. The set of 
equilibria for $\chi = 0$ coincides with the set of sequential equilibria.
Moreover, Figure \ref{cse} shows that as $\chi$ increases,
the education level of the type $\theta_H$ worker decreases under the weak set order, demonstrating that cursedness improves efficiency, since education does not enhance the
worker’s productivity.
As $\chi$ approaches 1, the only equilibrium is the most efficient pooling equilibrium,
where both types choose $e = 0$.

Lastly, it is worth noting that separating and pooling $\chi$-CSE are not the only types of equilibria.
Similar to the standard equilibrium analysis, hybrid $\chi$-CSE also exist,
where one type plays a pure strategy, while the other type mixes.
For instance, there exist hybrid $\chi$-CSE in which the type $\theta_H$ worker chooses
some $e_H$, while the type $\theta_L$ worker mixes between $e_L = e_H$ (with probability $q$)
and $e_L = 0$ (with probability $1 - q$). This strategy profile can be supported
as a hybrid $\chi$-CSE if
$$c(e_H| \theta_L) = (1-\chi)\left[\frac{p}{p+(1-p)q} \right](\theta_H - \theta_L). $$

\section{Cursed Intuitive Criterion}
\label{sec:cursed_intu_criter}

Aside from the fully cursed case ($\chi=1$), multiple separating and pooling 
$\chi$-CSE exist for any $\chi<1$. To sharpen our predictions and derive testable implications, we propose a belief-based 
refinement---the \emph{$\chi$-cursed 
intuitive criterion}---in the spirit of \cite{cho1987signaling}.
Our $\chi$-cursed intuitive criterion is defined 
for general signaling games. Therefore, we now briefly turn to the general framework before returning to Spence's model.

For any belief over types $\mu \in \Delta(\Theta)$ and message $m\in M$, 
we define the best response correspondence of the receiver as
$$BR(\mu, m) = \argmax_{\sigma_2 \in 
\Delta(A)} \sum_{\theta\in \Theta}
u_2(\theta, m, \sigma_2(m)) \cdot \mu(\theta) ,$$
and for any $\Xi \subset \Delta(\Theta)$, we define 
$$BR(\Xi, m) = \bigcup_{\mu\in \Xi} BR(\mu, m).$$
Furthermore, for any $\chi$-CSE assessment $(\mu, \sigma)$ with the 
associated equilibrium payoffs $u_{1}^{\chi*}(\theta)$
for the type $\theta$ sender, and for each 
off-path message $m$, we define 
$$T^\chi(m) = \left\{\theta\in \Theta: u_{1}^{\chi*}(\theta) >
\max_{\sigma_{2}\in BR(\Delta(\Theta), m)} 
u_1(\theta, m, \sigma_2(m)) \right\},$$
and let 
$$\Delta^\chi(m) = \left\{\mu\in \Delta(\Theta): 
\mu(\theta) = \chi F(\theta) \;\;\;\; \mbox{if} \;\;
\theta\in T^\chi(m) \right\}.$$

\begin{definition}\label{definition_cursed_intuitive_criterion}

A $\chi$-CSE assessment of a signaling game $(\mu, \sigma)$ satisfies the 
$\chi$-cursed intuitive criterion if, for any off-path message $m$, 
$\mu(\cdot|m)\in \Delta^\chi(m)$.
\end{definition}

$T^\chi(m)$ is the set of sender types who cannot possibly
benefit from sending the off-path message $m$, 
relative to their equilibrium 
payoff. In the same spirit as the intuitive criterion, 
upon receiving this off-path message $m$,
the receiver should assign the lowest possible 
probability weights to these types.
In the standard intuitive criterion, the receiver assigns 
zero weight to these types. However, due to the requirement of 
$\chi$-consistency, in $\chi$-CSE, the receiver's belief 
for each type $\theta$ is bounded below by 
$\chi F(\theta)$, regardless of the message received.
Accordingly, our $\chi$-cursed intuitive criterion requires the 
receiver to assign the lowest probabilities that satisfy 
$\chi$-consistency to the types in 
$T^\chi(m)$.

\begin{remark}
If there are no off-path events or if $T^\chi(m) = \emptyset$, then the $\chi$-CSE survives $\chi$-cursed
intuitive criterion. Moreover, two extreme cases are particularly noteworthy:   
\begin{itemize}
    \item When $\chi= 0$, our $\chi$-cursed intuitive criterion 
    reduces to the standard intuitive criterion. 
    \item When $\chi=1$, the fully cursed sequential equilibrium 
    always satisfies the cursed intuitive criterion. This is 
    because a fully cursed receiver's belief about the sender's 
    type coincides 
    with the prior distribution, i.e., $\mu^{\chi=1}(\theta | m) 
    = F(\theta)$ for all $\theta\in \Theta$ and all $m\in M$ by  
    $\chi$-cursed Bayes' rule. Consequently, this belief belongs to $\Delta^{\chi=1}(m)$.
\end{itemize}
\end{remark}

From the definition of the $\chi$-cursed intuitive criterion,
we can see that different values of $\chi$ impose different restrictions
on off-path beliefs, making the selection of the cursed 
intuitive criterion, in principle, highly sensitive to $\chi$.
As a result, despite their similar construction, the cursed intuitive 
criterion and the standard intuitive criterion are distinct refinement criteria. 
In Supplemental Appendix B, we provide an illustrative example of 
a modified beer-quiche game demonstrating that, in general, the two 
criteria may select different equilibria.

This prompts the question: which CSE survives the cursed intuitive criterion in the job market signaling game?
In Proposition \ref{prop:job_market_selection}, we prove that, similar to the 
selection under the standard intuitive criterion, the 
$\chi$-cursed intuitive criterion uniquely selects the most efficient separating 
$\chi$-CSE \emph{for any} $\chi$, except in the case of $\chi=1$, where the
unique CSE is the pooling equilibrium with $e_H = e_L = 0$.

\begin{proposition}\label{prop:job_market_selection}
For any $\chi\in[0,1]$, there is a unique $\chi$-CSE that survives the $\chi$-cursed 
intuitive criterion 
where $e_H = c^{-1}((1-\chi)(\theta_H - \theta_L)|\theta_L)$ and $e_L=0$.
\end{proposition}

This equilibrium outcome, known as the Riley outcome, 
requires the least inefficient signaling and therefore Pareto-dominates
all separating equilibria, as illustrated in the right panel of Figure \ref{cse}. In this sense, Proposition~\ref{prop:job_market_selection} demonstrates that the Riley outcome is robust to partial cursedness, as it survives the cursed intuitive criterion even when the receiver partially neglects the dependence of the sender's message on their private type.

However, a careful examination of the argument reveals that the \emph{cardinality
of the message space} plays a subtle yet crucial role in the existence of a 
separating $\chi$-CSE for any $\chi < 1$.
In the canonical setting, with a continuum of education levels, the high 
type worker can always choose a small but \emph{positive} education level to signal
their type, regardless of how cursed the players are.
In contrast, when the message space is discrete, a highly cursed worker can no 
longer choose an arbitrarily small amount of education, leading to a breakdown in information transmission.\footnote{An interesting theoretical question is whether $\chi$-CSE predictions converge to the continuous-message-space benchmark as the number of education choices increases. In general, the answer is sensitive to the structure of the message space. Nonetheless, we conjecture that under suitable conditions, the range of $\chi$ supporting a separating $\chi$-CSE expands monotonically with the number of available messages, converging to the full range $\chi \in [0,1)$
as the message space becomes arbitrarily rich. In Supplemental Appendix B, we extend the 
binary-choice setting of \cite{kubler2008job} to a trinary-choice setting and show that the 
range of $\chi$ supporting a separating $\chi$-CSE indeed expands relative to the binary case, consistent with this conjecture. We thank an anonymous referee for this 
observation.}

We regard the dependence of $\chi$-CSE predictions on the message-space cardinality as an important feature of the framework, rather than a limitation, for two reasons. First, it speaks to the ``Message Space Puzzle'' in information economics: whether and when the size of the message space affects information transmission in equilibrium. While a smaller message space leads to inefficiency in some environments \citep{crawford1982strategic, heumann2020cardinality} and is payoff-irrelevant in others (e.g., \citealt{myerson1982optimal, kamenica2011bayesian}), $\chi$-CSE highlights an interplay between cursedness and richness of the message space: information transmission can be sustained in a coarse message space only when players are not too cursed.\footnote{In other words, if the firm were able to choose the size of the message space, a cursed firm would strictly \emph{prefer} a richer message space in order to preserve information transmission. This observation is consistent with the recent experimental findings of \cite{je2024signal}, who show that individuals favor signals drawn from larger spaces even in the absence of any instrumental benefit.}
Second, this dependence yields a sharp, experimentally testable prediction that distinguishes cursed intuitive criterion from the standard intuitive criterion: with a sufficiently coarse message space, the cursed intuitive criterion predicts that separation breaks down, whereas the standard intuitive criterion does not. We examine this prediction in the next section by revisiting the job market signaling experiment of \cite{kubler2008job}, which features two worker types and a binary education choice.

\section{Experimental Evidence from \cite{kubler2008job}}
\label{sec:experimental_evidence}

In the experimental job market signaling game of \cite{kubler2008job}, 
the productivity of types $\theta_H$ and $\theta_L$ are 50 and 10, respectively. Each type is drawn with equal probability. The worker's education 
choice is binary, $s \in \{0,1\}$, with education costs of 9 for type 
$\theta_H$ and 45 for type $\theta_L$. As shown in \cite{kubler2008job}, even with only two messages, the intuitive criterion 
selects the separating equilibrium. In contrast, the set of $\chi$-CSE 
that survives the $\chi$-cursed intuitive criterion varies with different
values of $\chi$, illustrating how a restrictive message space can obstruct 
information transmission. This highlights a fundamental difference between the 
cursed intuitive criterion and the
standard intuitive criterion as refinement concepts.

\begin{figure}[H]
    \centering
    \includegraphics[width=0.9\linewidth]{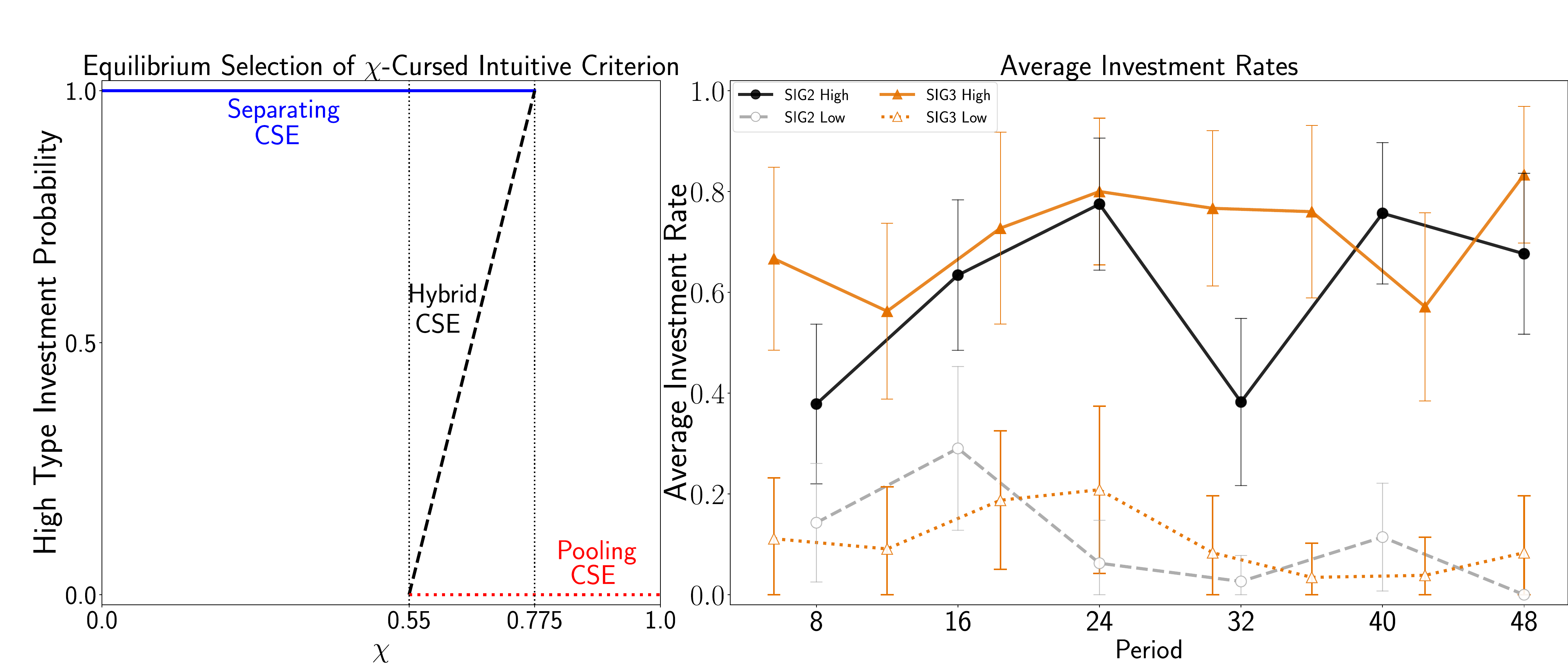}
    \caption{(Left) The set of $\chi$-CSE that survives the $\chi$-cursed intuitive criterion.
    (Right) Average investment rates with 95\% CIs in SIG2 and SIG3 from \cite{kubler2008job} across different blocks.}
    \label{fig:cursed_intuitive_experiment_analysis}
\end{figure}

The formal analysis can be found in Supplemental Appendix B. 
For illustration purposes, we plot the set of $\chi$-CSE that survives the $\chi$-cursed
intuitive criterion in the left panel of Figure
\ref{fig:cursed_intuitive_experiment_analysis}. Since the low type worker never 
invests in education in any equilibrium, we focus on changes in the high type worker's choice
as $\chi$ varies.
From the figure, we observe that our $\chi$-cursed intuitive criterion uniquely selects the 
separating equilibrium only for sufficiently small $\chi$ (i.e., for $\chi < 0.55$). 
Yet for intermediate values of $\chi \in \left(0.55, 0.775\right)$,
the criterion is too weak to eliminate any equilibrium: separating, pooling, 
and hybrid CSE all survive. Finally, for $\chi \geq 0.775$, the pooling 
equilibrium, in which neither type invests in education and private information is not transmitted, 
is the only equilibrium that survives the cursed intuitive criterion. In short, when the 
message space becomes too coarse, cursedness can deteriorate rather than improve
the efficiency of information transmission.

To assess whether the experimental data aligns with the predictions of the cursed 
intuitive criterion, we reanalyze the signaling treatment data from their experiment. 
Instead of assuming that the firm has a quadratic loss payoff function, the experiment considers
a setting with multiple firms, where the firm that offers the highest 
wage employs the worker. If a firm hires the worker, the firm's payoff is given by the difference between 
the worker's true productivity and the wage.

The experiment includes two firms in the SIG2 treatment and three 
firms in the SIG3 treatment.\footnote{Their experiment also includes two 
screening treatments in which firms first offer wages, then the worker decides 
whether to invest in education. We focus
on the two signaling treatments, given their relevance to our theory. A detailed description of the experiment 
and its instructions can be found in \cite{kubler2008job}.}
Each SIG2 session consists of 9 subjects, while each SIG3 session consists of 12 subjects. 
There are three sessions per treatment, resulting 
in a total of 27 subjects in SIG2 and 36 subjects in SIG3. All sessions 
last 48 periods. In each period, the groups are randomly rematched,
and workers' types are randomly redrawn.
In the SIG2 treatment, the 48 periods are divided into six blocks of eight
consecutive periods each. Roles remain fixed within each block, with all 
subjects playing as workers for two blocks and as firms for four blocks. 
Similarly, in the SIG3 treatment, the 48 periods are divided into eight 
blocks of six consecutive periods each, with all subjects playing as
workers for two blocks and as firms for six blocks.

The right panel of Figure \ref{fig:cursed_intuitive_experiment_analysis} 
plots the average investment 
rates in each block for each type in both treatments. From the figure, 
we can see that as the experiment progresses, low type workers in both treatments
realize that it is optimal not to invest in education. In the last block of
the SIG2 treatment, none of the low type workers choose to invest. Similarly, 
in the last block of the SIG3 treatment, the average investment rate of 
low type workers is only 8.3\% (two-tailed t-test p-value $= 0.162$).

On the contrary, even with repetitions, high type workers' behavior 
does not converge to the prediction of the standard intuitive criterion. 
In the SIG2 treatment, only 37.8\% of high type workers invest 
in education in the first block, which is significantly lower 
than the prediction of the intuitive criterion (two-tailed t-test p-value 
$<0.001$). The average investment rates fluctuate 
in the subsequent blocks. In the last block, the average investment rate is 67.6\%, 
which remains significantly lower than 1 (two-tailed t-test p-value $<0.001$).
A similar pattern emerges in the SIG3 treatment. Although high type workers 
are more likely to invest in education in the SIG3 treatment, the average
investment rate remains significantly lower than the prediction of the intuitive 
criterion in the last block (two-tailed t-test p-value $=0.023$).\footnote{See Supplemental Appendix C for a block-by-block analysis.}

In summary, the experimental results of \cite{kubler2008job} suggest a violation of the intuitive criterion in the job market signaling game: with repetition, low type workers' choices converge to the theoretical prediction, while high type workers' choices deviate from it in a manner consistent with our cursed intuitive criterion. We view this alignment as suggestive rather than definitive evidence, since cognitive load and task complexity may differ across types, and high type workers' deviation from full investment could reflect other behavioral biases rather than cursedness. To rule out confounds that do not depend on message space richness, a more direct test would experimentally manipulate the number of available education levels: our theory predicts that the bite of the cursed intuitive criterion depends on this margin.


\section{Application: Wage Compression in Labor Markets}
\label{sec:conclusion}

Beyond this experimental setting, the underlying mechanism we 
identify---cursedness narrowing the wage gap firms offer across 
education levels---has a direct empirical counterpart outside the laboratory: \emph{wage compression}, the narrowing wage gap between workers of different experience or skill levels. This has been observed across countries \citep{campbell1997reasons, mourre2005wage} and industries \citep{frank1984workers}, and has recurred historically---most notably in the Great Compression of the 1940s \citep{goldin1992great} and again post-pandemic \citep{autor2023unexpected}. We illustrate wage compression below in the canonical job market signaling game with a continuum of types.\footnote{This 
setting follows Exercise 13.C.4 in \cite{mas1995microeconomic}. Although CSE is formally defined only for games with finite types in \cite{fong2023cursed}, 
the following analysis demonstrates how CSE can
be naturally extended to games with a continuum of types.}


Suppose the worker's productivity type $\theta$ is distributed over a compact set 
$[\underline{\theta}, \bar{\theta}]$, according to a distribution with 
strictly positive density function $f(\theta)$, with $f(\theta)>0$ for all $\theta
\in[\underline{\theta}, \bar{\theta}]$ and $0<\underline{\theta}< 
\bar{\theta} $.
To derive a closed-form solution, we assume that the cost function 
for the type $\theta$ worker with an education level $e$ is $c(e|\theta)= \frac{e^2}{\theta}$.
Under this setting, Proposition 
\ref{prop:continuum_separating} characterizes the unique separating $\chi$-CSE
for $\chi<1$, and highlights that no information is transmitted when $\chi=1$.

\begin{proposition}\label{prop:continuum_separating}
In the job market signaling game with a continuum of types, 
\begin{itemize}
    \item[1.] for any $\chi<1$, there exists a unique separating $\chi$-CSE where 
    the education level of type $\theta$ is given by 
    $$e^\chi(\theta) = \sqrt{\frac{1}{2}(1-\chi)(\theta^2 - \underline{\theta}^2)}$$
    and upon observing $e^\chi(\theta)$, the firm offers a wage of $w(e^\chi(\theta)) = 
    \chi\mathbb{E}[\theta] + (1-\chi)\theta$.
    \item[2.] when $\chi=1$, the unique CSE is the pooling equilibrium where 
    $e^{\chi=1}(\theta)=0$ for any $\theta$ and the firm offers a
    wage of $\mathbb{E}[\theta]$ upon observing any $e$.
\end{itemize}
\end{proposition}

\begin{figure}[H]
    \centering
    \includegraphics[width=0.9\linewidth]{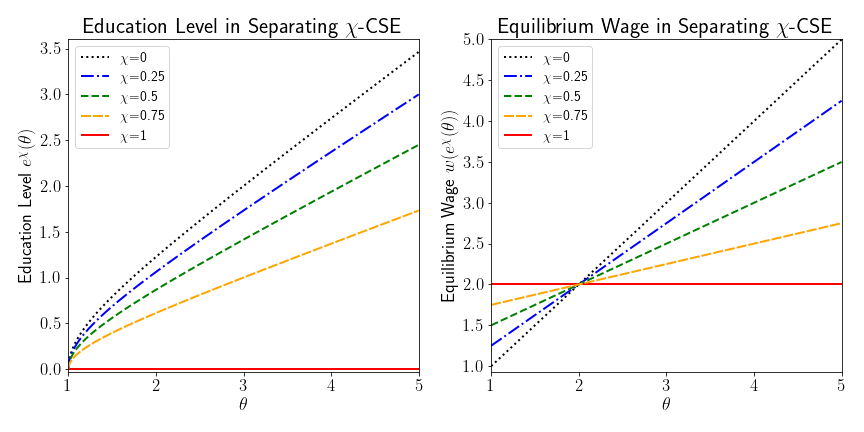}
    \caption{(Left) The education level $e^\chi(\theta)$ and (Right) the 
    equilibrium wage $w(e^\chi(\theta))$ in the separating $\chi$-CSE with 
    $\underline{\theta}=1$ and $\mathbb{E}[\theta] = 2$.}
    \label{fig:continuum_sep}
\end{figure}

The key insight from the two-type model (with a continuum of messages) is that, regardless 
of how cursed the players are, a separating $\chi$-CSE exists as 
long as $\chi < 1$. Proposition \ref{prop:continuum_separating} confirms 
that this result holds even when the model is extended to accommodate a 
continuum of types. To illustrate the effect of cursedness, we plot the 
equilibrium education level $e^\chi(\theta)$ in the left panel of
Figure \ref{fig:continuum_sep}. 
The figure shows that as the firm becomes more cursed,
the worker \emph{deflates} the value of education, yet education remains 
informative as long as $\chi < 1$.
Consequently, as $\chi \to 1$, information transmission approaches maximum 
efficiency, with full revelation attained at almost negligible cost. However, when 
$\chi = 1$, information transmission breaks down.\footnote{In addition to 
the unique separating $\chi$-CSE, there exists a family of pooling 
$\chi$-CSE: $e^\chi(\theta)=\bar{e}$ 
for all $\theta$ can be supported as a pooling $\chi$-CSE if 
    $\bar{e}\leq \sqrt{(1-\chi)\left[\underline{\theta}
    \mathbb{E}[\theta] - \underline{\theta}^2\right]}$
which guarantees incentive compatibility of the lowest type $\underline{\theta}$
(and hence all higher types).}

Moreover, in the right panel of Figure \ref{fig:continuum_sep}, 
we plot the equilibrium wage $w(e^\chi(\theta))$ for each type. We find that,
as players become more cursed, the wage scheme pivots around 
the average type $\mathbb{E}[\theta]$, and the slope of the wage function flattens.
Specifically, as cursedness increases, wages converge toward 
the mean: workers with below-average productivity receive higher wages,
while those with above-average productivity receive lower wages.
This illustrates how cursedness can lead to wage compression in the job market signaling game.

An extensive literature explores the origins of wage compression, including 
unionization and collective bargaining \citep{freeman1982union,farber2021unions, mogstad2025income, baker2026impact}, fairness concern \citep{akerlof1990fair, fehr2025social}, over- and under-confidence \citep{santos2012labor}, and uncertainty about productivity \citep{gross2015merit, breza2018morale}. Our result complements this literature, suggesting cursedness may also contribute to this phenomenon.

Among the explanations above, uncertainty about ability sits closest to ours. \cite{gross2015merit} show experimentally that managers compress wages between perceived high- and low-ability workers more when their uncertainty about worker ability is greater. Our channel produces compression without any such uncertainty: the firm observes education perfectly, but cursedness makes the firm discount how strongly that signal reveals type. Cursedness, in this sense, offers a theoretical possibility for wage compression that does not require appeal to either social preferences or institutional rigidity.\footnote{As documented in \cite{autor2023unexpected}, wage dispersion moved in both directions in recent decades. We hypothesize that cursedness is one contributing channel among several, rather than the dominant force at any given time. This view is consistent with \cite{camerer2022using}, who document the effects of gender, experience, and stakes on wage bargaining, and with \cite{capra2026virtual}, who find attenuation of strategic misperception under salient fairness norms.}


\bibliographystyle{ecta}

\bibliography{reference}

\newpage
\appendix

\setcounter{proposition}{0}
\renewcommand{\theproposition}{S.\arabic{proposition}}

\setcounter{figure}{0}
\renewcommand{\thefigure}{S.\arabic{figure}}

\setcounter{table}{0}
\renewcommand{\thetable}{S.\arabic{table}}

\setcounter{section}{0}
\renewcommand{\thesection}{Appendix \Alph{section}}

\section{Omitted Proofs}

\subsection*{Proof of Proposition \ref{prop:sep_cse_ic}}

To simplify the exposition, let $e_\theta$
denote the education choice of the type $\theta$ worker, 
and let $\mu^\chi(e) \equiv \mu^\chi(\theta_H | e)$ denote the firm's belief 
that the worker is of the high type given the observed education level $e$ 
in a $\chi$-CSE. With this notation, we first observe that, for any posterior belief 
$\mu^\chi(e)$, due to the quadratic payoff function, the firm will offer a wage of
\begin{align*}
w(\mu^\chi(e)) =  \mu^\chi(e)\theta_H 
    + (1 - \mu^\chi(e)) \theta_L.
\end{align*}
In any separating $\chi$-CSE where $e_H \neq e_L$, 
upon observing $e_H$ and $e_L$ (on-path events), 
the firm will update their beliefs using the 
$\chi$-cursed Bayes' rule as follows:
\begin{align*}
\mu^\chi(e_H) = \chi p + (1-\chi) \;\;\;\;
\mbox{and} \;\;\;\; \mu^\chi(e_L) = \chi p.
\end{align*}
Therefore, on the equilibrium path, the firm will offer  
$w^\chi_H \equiv w(\mu^\chi(e_H))$ upon observing $e_H$
and $w^\chi_L \equiv w(\mu^\chi(e_L))$ upon observing $e_L$ such that 
\begin{align*}
w^\chi_H = \theta_H - (1-p)\chi(\theta_H - \theta_L) \;\;\;\;
\mbox{and} \;\;\;\; w^\chi_L = \theta_L + p\chi(\theta_H - \theta_L).
\end{align*}
Due to the requirements of $\chi$-consistency and 
sequential rationality, in any 
separating $\chi$-CSE, the equilibrium wage, conditional
on observing any education level, will lie 
within the interval $[w_L^\chi, w_H^\chi]$.

With this observation, we can show by contradiction that 
in any separating $\chi$-CSE, the type 
$\theta_L$ worker will choose 
$e_L=0$. Suppose this is not true---that the type 
$\theta_L$ worker instead chooses some 
$e'>0$. In this case, deviating to 
$e=0$ would be profitable for the type 
$\theta_L$  worker, as their wage would still 
be at least $w_L^\chi$
without paying any cost, yielding a 
contradiction.

Moreover, to support a separating $\chi$-CSE, the IC conditions
for both types must be 
satisfied. For the type $\theta_H$ worker, 
the IC condition is
\begin{align}\label{eq:IC_sep_high}
w^\chi_L - c(0|\theta_H) \leq w^\chi_H - c(e_H| \theta_H) 
\iff c(e_H|\theta_H)\leq(1-\chi)(\theta_H-\theta_L),
\tag{IC-$\theta_H$}
\end{align}
and for the type $\theta_L$ worker, the IC condition is
\begin{align}\label{eq:IC_sep_low}
w^\chi_L - c(0|\theta_L) \geq w^\chi_H - c(e_H| \theta_L) \iff 
c(e_H|\theta_L)\geq(1-\chi)(\theta_H-\theta_L). 
\tag{IC-$\theta_L$}
\end{align}
Combining the two IC conditions, we obtain that in any
separating $\chi$-CSE, 
$$c(e_H|\theta_L)\geq(1-\chi)(\theta_H-\theta_L)\geq c(e_H|\theta_H).$$

Finally, for any off-path event where the worker chooses 
$e\not\in \{e_H, e_L \}$, we specify the 
firms' belief as $\mu^\chi(e) = \chi p$, 
which is the lowest belief allowed by 
Proposition \ref{prop:chi_dampen_update}. 
In this case, neither type of worker has a profitable 
deviation. $\blacksquare$

\subsection*{Proof of Corollary \ref{coro:monotone_separating}}

Due to the strict 
monotonicity of the cost function, $\mathcal{S}^\chi$
can be characterized as 
$$e_H\in \mathcal{S}^\chi \iff 
c^{-1}((1-\chi)(\theta_H - \theta_L)|\theta_L)\leq e_H \leq
c^{-1}((1-\chi)(\theta_H - \theta_L)|\theta_H).$$
For any $\chi$ and $\chi'$ such that $\chi \geq \chi'$, we know $(1-\chi)(\theta_H - \theta_L)
\leq (1-\chi')(\theta_H - \theta_L) $, 
implying that $c^{-1}((1-\chi)(\theta_H - \theta_L)|\theta) \leq 
c^{-1}((1-\chi')(\theta_H - \theta_L)|\theta)$ for $\theta\in \{\theta_L, \theta_H \}$.
That is, both the upper bound and the lower bound decrease with $\chi$. 
Therefore, for any $e_H \in \mathcal{S}^\chi$, $e_H \leq c^{-1}((1-\chi')(\theta_H - \theta_L)|\theta_H)
\in \mathcal{S}^{\chi'}$, and for any $e_H' \in \mathcal{S}^{\chi'}$, 
$e_H' \geq c^{-1}((1-\chi)(\theta_H - \theta_L)|\theta_L) \in \mathcal{S}^{\chi}$. 
This proves that $\mathcal{S}^{\chi'} \geq_{ws} \mathcal{S}^\chi$. $\blacksquare$

\subsection*{Proof of Proposition \ref{prop:pool_cse_ic}}

In any pooling $\chi$-CSE where $e_L = e_H = e^*$, upon
observing $e^*$ (the on-path event), the firm's belief remains the prior, 
$\mu^\chi(e^*) = p$, by $\chi$-cursed Bayes' rule. Thus, on the equilibrium path,
the firm will offer $w^{\chi*} = p\theta_H + (1 - p) \theta_L.$

Similar to the proof of Proposition \ref{prop:sep_cse_ic}, for any off-path event 
where the worker chooses $e\neq e^*$, we specify the firm's belief as $\mu^\chi(e)=\chi p$ and the firm offers $w^{\chi}_0 = \chi p \theta_H + (1-\chi p)\theta_L$.
Lastly, to support the pooling $\chi$-CSE, the following conditions 
must be satisfied to ensure that neither type has an incentive
to deviate by choosing $e = 0$. That is, for the type $\theta_H$ worker,
\begin{align}\label{eq:IC_pool_high}
w^{\chi}_0 - c(0|\theta_H) \leq w^{\chi*} - c(e^*| \theta_H)
\iff c(e^*|\theta_H)\leq(1-\chi)p(\theta_H-\theta_L) 
\tag{pooling-$\theta_H$}
\end{align}
and for the type $\theta_L$ worker, 
\begin{align}\label{eq:IC_pool_low}
w^{\chi}_0 - c(0|\theta_L) \leq w^{\chi*} - c(e^*| \theta_L) 
\iff c(e^*|\theta_L)\leq(1-\chi)p(\theta_H-\theta_L) 
\tag{pooling-$\theta_L$}
\end{align}
Because $c(e^*|\theta_L) \geq c(e^*|\theta_H)$, we know the condition (\ref{eq:IC_pool_high})
is never binding. Therefore, we conclude that in any pooling $\chi$-CSE where 
$e_L = e_H = e^*$, 
$c(e^*| \theta_L) \leq (1-\chi)p(\theta_H - \theta_L). $ This completes 
the proof. $\blacksquare$

\subsection*{Proof of Corollary \ref{coro:monotone_pooling}}

Due to the strict 
monotonicity of the cost function, $\mathcal{P}^\chi$
can be characterized as 
$$e\in \mathcal{P}^\chi \iff 
0\leq e \leq
c^{-1}((1-\chi)p(\theta_H - \theta_L)|\theta_L).$$
For any $\chi$ and $\chi'$ such that $\chi \geq \chi'$, we know $(1-\chi)p(\theta_H - \theta_L)
\leq (1-\chi')p(\theta_H - \theta_L) $, 
implying that $c^{-1}((1-\chi)p(\theta_H - \theta_L)|\theta_L) \leq 
c^{-1}((1-\chi')p(\theta_H - \theta_L)|\theta_L)$, 
i.e., the upper bound decreases with $\chi$. 
Therefore, for any $e \in \mathcal{P}^\chi$, $e \leq c^{-1}((1-\chi')p(\theta_H - \theta_L)|\theta_L)
\in \mathcal{P}^{\chi'}$, and for any $e \in \mathcal{P}^{\chi'}$, 
$e \geq 0 \in \mathcal{P}^{\chi}$. 
Hence, $\mathcal{P}^{\chi'} \geq_{ws} \mathcal{P}^\chi$. 
Finally, when $\chi=1$, $\mathcal{P}^{\chi=1} = \{0 \}$. Combining with 
Proposition \ref{prop:sep_cse_ic}, we conclude that the 
unique fully cursed sequential equilibrium 
$(\chi=1)$ is the pooling equilibrium where $e_L = e_H = 0$.
$\blacksquare$

\subsection*{Proof of Proposition \ref{prop:job_market_selection}}

We first note that when $\chi=1$, the firm's belief 
about the worker's type remains the prior, i.e., $\mu^{\chi=1}(e)=p$ for any $e$. 
Since the prior distribution belongs to $\Delta^{\chi=1}(e)$ for any $e$, 
the fully cursed sequential equilibrium survives the cursed 
intuitive criterion for $\chi=1$. Therefore, in the rest of the proof,
which consists of two steps, we consider the case where 
$\chi\in[0,1)$. In the first step,
we show that no pooling or hybrid equilibrium
survives the $\chi$-cursed intuitive criterion. In the second step, 
we prove that the unique $\chi$-CSE that survives $\chi$-cursed 
intuitive criterion is the most efficient separating $\chi$-CSE.

\bigskip

\noindent\textbf{Step 1: No pooling or hybrid equilibrium
survives $\chi$-cursed intuitive criterion}

\medskip

First, suppose that in equilibrium both types $\theta_H$ and $\theta_L$ choose $e$ with positive probability, where $\sigma(e|\theta_H)=h>0$ and 
$\sigma(e|\theta_L)=l>0$. 
In this case, upon observing $e$, by Proposition~\ref{prop:curse_bayes_rule},
the firm's belief that the worker is of type $\theta_H$ is
\begin{align}\label{inequ:general_cic_proof}
\mu^\chi(e) = \chi p + (1-\chi)\left[\frac{ph}{ph + (1-p)l} \right] < 1 - (1-p)\chi. \tag{S.1} 
\end{align}
Note that this inequality holds for any education level $e$. It will be useful in the proof below, as it provides a strict upper bound on the firm's belief that the worker is of the high type in any pooling or hybrid $\chi$-CSE.

Second, we show by contradiction that no pooling or hybrid equilibrium survives the $\chi$-cursed intuitive criterion. Suppose not; there exists an equilibrium where both types $\theta_H$ and $\theta_L$ choose some $\bar{e}$ with positive probability, and the expected payoff to type~$\theta$ from choosing $\bar{e}$ is
$\mu^\chi(\bar{e})\theta_H + (1-\mu^\chi(\bar{e}))\theta_L - c(\bar{e}|\theta)$.
Let $e'>\bar{e}$ be the education level such that 
$$[1-(1-p)\chi]\theta_H+(1-p)\chi\theta_L-c(e'|\theta_L) = \mu^\chi(\bar{e})\theta_H +
(1-\mu^\chi(\bar{e}))\theta_L - c(\bar{e}|\theta_L).$$
Such an $e'$ exists because inequality (\ref{inequ:general_cic_proof}) ensures that 
$\mu^\chi(\bar{e}) < 1 - (1-p)\chi$.

Now consider an $e'' > e'$ that is close to $e'$. The type $\theta_L$ worker
does not have an incentive to deviate to $e''$ because 
$$\mu^\chi(\bar{e})\theta_H +
(1-\mu^\chi(\bar{e}))\theta_L - c(\bar{e}|\theta_L) > 
[1-(1-p)\chi]\theta_H+(1-p)\chi\theta_L-c(e''|\theta_L).$$
In contrast, it is possible for the type $\theta_H$ worker to deviate to $e''$ as
\begin{align}\label{inequ:curesed_ict_pool}
\mu^\chi(\bar{e})\theta_H +
(1-\mu^\chi(\bar{e}))\theta_L - c(\bar{e}|\theta_H) < 
[1-(1-p)\chi]\theta_H+(1-p)\chi\theta_L-c(e''|\theta_H)& \tag{S.2} \\
\iff c(e''| \theta_H) - c(\bar{e}|\theta_H) <  c(e'| \theta_L) - c(\bar{e}|\theta_L)&, \notag
\end{align}
and the last inequality holds due to the assumption that type 
$\theta_H$ has a strictly lower marginal cost than type
$\theta_L$. Consequently, $T^\chi(e'') = \{\theta_L \}$ and the $\chi$-cursed intuitive 
criterion requires $\mu^\chi(e'') = 1 - (1-p)\chi$.
By inequality (\ref{inequ:curesed_ict_pool}), we know that type $\theta_H$ will 
deviate, implying that the equilibrium fails the $\chi$-cursed intuitive criterion, 
yielding a contradiction.

\bigskip

\noindent\textbf{Step 2: Most efficient separating $\chi$-CSE is the 
unique selection}

\medskip

We let $\underline{e}_H^\chi \equiv c^{-1}((1-\chi)(\theta_H - \theta_L)|\theta_L)$ be the lowest 
education level that the type $\theta_H$ worker can attain in a separating $\chi$-CSE. Now, we prove that any separating $\chi$-CSE where type $\theta_H$ chooses $e_H > \underline{e}_H^\chi $
and type $\theta_L$ chooses 0 violates the $\chi$-cursed intuitive criterion. 

Consider a separating $\chi$-CSE where type $\theta_H$ chooses $e_H > \underline{e}_H^\chi $
and type $\theta_L$ chooses 0. Fix $e' \in (\underline{e}_H^\chi, e_H)$. In this case, since 
(\ref{eq:IC_sep_low}) binds at $\underline{e}_H^\chi$, we conclude that type $\theta_L$ will 
not deviate to choose $e'$ as 
\begin{align*}
[1-(1-p)\chi]\theta_H + (1-p)\chi\theta_L - c(e'|\theta_L) 
<& [1-(1-p)\chi]\theta_H + (1-p)\chi\theta_L - c(\underline{e}_H^\chi|\theta_L)    \\
=& \chi p \theta_H + (1-\chi p)\theta_L.
\end{align*}
Yet, it is possible for type $\theta_H$ to deviate to $e'$,
because they may believe that $e'$ yields the same wage as $e_H$
but at a lower cost. That is, $T^\chi(e') = \{\theta_L \} $ and the $\chi$-cursed intuitive 
criterion requires the belief upon observing $e'$ to be $\mu^\chi(e') = 1 - (1-p)\chi$.
As a result, type $\theta_H$ will deviate from $e_H$ to $e'$, because the wage remains the 
same, while the cost is strictly lower. This proves that any separating 
$\chi$-CSE with 
$e_H >\underline{e}_H^\chi$ violates the 
$\chi$-cursed intuitive criterion. $\blacksquare$

\subsection*{Proof of Proposition \ref{prop:continuum_separating}}

First, in a separating $\chi$-CSE, all types choose different education levels.
Let $e^\chi(\theta)$ denote the education level chosen by type $\theta$
in a separating $\chi$-CSE. For any 
$\theta \neq \theta'$, we have $e^\chi(\theta)\neq e^\chi(\theta')$. Therefore, upon observing $e^\chi(\bar{\theta})$, 
$\chi$-cursed Bayes' rule implies that the firm's belief density function $\mu^\chi(\theta|e^\chi(\bar{\theta}))$ 
is given by:
$$\mu^\chi(\theta | e^\chi(\bar{\theta})) = \chi f(\theta) + (1-\chi)\mathbbm{1}\{\theta = \bar{\theta} \}. $$
Consequently, the firm will offer a wage of
$w(e^\chi(\theta))= \chi \mathbb{E}[\theta] + (1-\chi)\theta$.

Second, consider the case where $\chi<1$.
Since the worker maximizes the expected payoff, the first-order
condition is given by $w'(e^\chi(\theta)) = 2e^\chi(\theta)/\theta$ for all $\theta$.
Moreover, from $w(e^\chi(\theta))= \chi \mathbb{E}[\theta] + (1-\chi)\theta$, we can derive 
that $w'(e^\chi(\theta))\cdot e^\chi(\theta)' = 1- \chi >0 $, implying that
$e^\chi(\theta)$ is monotonically 
increasing in $\theta$. In this case, type $\underline{\theta}$ can be identified and hence 
$e^\chi(\underline{\theta})=0$.
As we substitute the equilibrium wage into the worker's first-order condition, 
it becomes a linear differential equation: 
$$w'(e^\chi(\theta)) = \frac{2(1-\chi)e^\chi(\theta)}{w(e^\chi(\theta))-\chi\mathbb{E}[\theta]}
\iff w(e^\chi(\theta))w'(e^\chi(\theta)) - \chi\mathbb{E}[\theta]w'(e^\chi(\theta)) = 2(1-\chi)e^\chi(\theta).$$
Integrating both sides, we obtain
\begin{align}\label{eq:general_ode}
\frac{w(e^\chi(\theta))^2}{2}-\chi \mathbb{E}[\theta]w(e^\chi(\theta))= (1-\chi) e^\chi(\theta)^2 + C, \;\; 
C\in\mathbb{R}.   \tag{S.3} 
\end{align}
Furthermore, because type $\underline{\theta}$ will choose $e^\chi(\underline{\theta})=0$ with an expected 
wage of $\chi\mathbb{E}[\theta] + (1-\chi)\underline{\theta}$, we can 
uniquely pin down the constant $C$ to be
\begin{align*}
\frac{w(0)^2}{2}-\chi \mathbb{E}[\theta]w(0)=C 
\iff &\frac{[\chi\mathbb{E}[\theta]+(1-\chi)\underline{\theta}]^2}{2}-\chi \mathbb{E}[\theta]\,[\chi\mathbb{E}[\theta]+(1-\chi)\underline{\theta}]=C \\
\iff &\frac{1}{2}(1-\chi)^2\underline{\theta}^2-\frac{1}{2} \chi^2 \mathbb{E}[\theta]^2 = C.
\end{align*}
We can then derive the equilibrium education function $e^\chi(\theta)$ by substituting $C$ into (\ref{eq:general_ode}):
\begin{align*}
& \frac{w(e^\chi(\theta))^2}{2}-\chi \mathbb{E}[\theta]w(e^\chi(\theta))= (1-\chi) e^\chi(\theta)^2 + \frac{1}{2}(1-\chi)^2\underline{\theta}^2-\frac{1}{2} \chi^2 \mathbb{E}[\theta]^2\\
\iff &\frac{1}{2}(1-\chi)^2 \theta^2-\frac{1}{2} \chi^2 \mathbb{E}[\theta]^2=  (1-\chi) e^\chi(\theta)^2 +\frac{1}{2}(1-\chi)^2\underline{\theta}^2-\frac{1}{2} \chi^2 \mathbb{E}[\theta]^2\\
\iff &e^\chi(\theta)=\sqrt{\frac{1}{2}(1-\chi)(\theta^2-\underline{\theta}^2)}.
\end{align*}

Lastly, we note that when $\chi=1$, the firm does not update their belief, meaning 
$\mu^{\chi=1}(\theta | e) = f(\theta)$ for any 
$e$ and $\theta$. Consequently, in equilibrium, the worker chooses 
$e^{\chi=1}(\theta)=0$ for all $\theta$, and the firm offers a wage of 
$w(e) = \mathbb{E}[\theta]$ upon observing any $e$. $\blacksquare$


\section{Additional Results}

\subsection*{Example: Cursed Intuitive Criterion $\neq$ Intuitive Criterion}

\begin{figure}[htbp!]
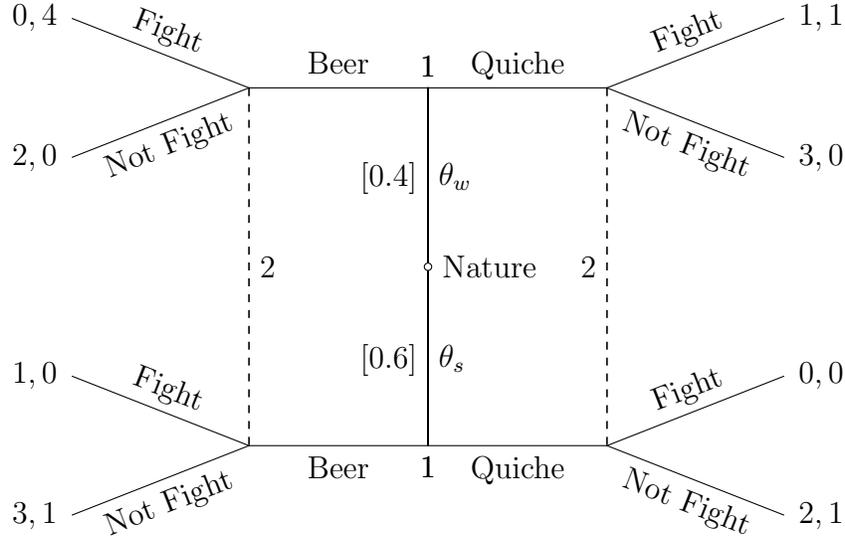

\centering
\begin{istgame}[scale=0.85]
\setistNewNodeStyle{solid node}[null node]
\setistmathTF001
\xtdistance{28mm}{22mm}
\istroot(0)[chance node]<0>{Nature}
  \istB<grow=90>{[0.4]}[l]{$\theta_w$}[r]
  \istB<grow=-90>{[0.6]}[l]{$\theta_s$}[r]
  \endist
\setistgrowdirection{west}
\istroot(a)(0-1)<90>{1}
  \istb{Beer}[a]
  \endist
\istroot(b)(0-2)<-90>{1}
  \istb{Beer}[b]
  \endist
\istroot(T2a)(a-1)
  \istb{Fight}[above,sloped]{0,4}
  \istb{Not Fight}[below,sloped]{2,0}
  \endist
\istroot(T2b)(b-1)
  \istb{Fight}[above,sloped]{1,0}
  \istb{Not Fight}[below,sloped]{3,1}
  \endist
\setistgrowdirection'{east}
\istroot(c)(0-1)<90>{1}
  \istb{Quiche}[a]
  \endist
\istroot(d)(0-2)<-90>{1}
  \istb{Quiche}[b]
  \endist
\istroot(T2c)(c-1)
  \istb{Fight}[above,sloped]{1,1}
  \istb{Not Fight}[below,sloped]{3,0}
  \endist
\istroot(T2d)(d-1)
  \istb{Fight}[above,sloped]{0,0}
  \istb{Not Fight}[below,sloped]{2,1}
  \endist
\xtInfoset[dashed](T2a)(T2b){2}[r]
\xtInfoset[dashed](T2c)(T2d){2}[l]
\end{istgame}
\caption{The modified beer-quiche game.}\label{beer_quiche}
\end{figure}

Consider the modified beer-quiche game with the 
game tree shown in Figure \ref{beer_quiche}. In this game, 
player 1 has two possible types, $\Theta = \{\theta_w, 
\theta_s\}$, with 
the prior probability of $\theta_w$ given by $F(\theta_w) = 0.4$.
Player 1 can choose from the set $M = \{\mbox{Beer}, \mbox{Quiche} \}$,
while player 2 can choose from the set
$A = \{\mbox{Fight}, \mbox{Not Fight}\}$.
In the following analysis, we use the four-tuple 
$[(\sigma_1(\mbox{Beer}|\theta_w),
\sigma_1(\mbox{Beer}|\theta_s)); (\sigma_2(\mbox{Fight}|\mbox{Beer}),
\sigma_2(\mbox{Fight}|\mbox{Quiche}))]$ to denote a 
behavioral strategy profile of this game.

There is a continuum of pooling equilibria in this game:
$[(0, 0); (q, 0)]$ where $q\geq 0.5$, yet none survives the standard
intuitive criterion.
The only sequential equilibrium that does is the semi-separating equilibrium
$[(3/8, 1); (1/2, 1)]$, where type $\theta_w$
mixes between Beer and Quiche, while type $\theta_s$
chooses Beer with probability 1.

In contrast, when $\chi=0.5$, 
the entire continuum of pooling equilibria: $[(0, 0); (q, 0)]$ where $q\geq 0.5$
survives the cursed intuitive criterion. Moreover, 
when $\chi>0.5$, the pooling equilibrium 
$[(0,0);(1, 0)]$, where both types of player 1 choose Quiche, 
is the unique pooling $\chi$-CSE surviving the $\chi$-cursed 
intuitive criterion.
In this equilibrium, type $\theta_w$ player 1 has already achieved
their maximum payoff, whereas type $\theta_s$ player 1 has not.
Therefore, $T^{\chi}(\mbox{Beer})= \{\theta_w \}$, and the off-path belief
must be $\mu^{\chi}(\theta_w | \mbox{Beer})= 0.4\chi \geq 0.2 \iff \chi\geq 0.5$, 
implying that the pooling equilibrium $[(0,0);(1, 0)]$ survives the 
$\chi$-cursed intuitive criterion for $\chi\geq 0.5$. 
$\square$

\subsection*{Formal Analysis of \cite{kubler2008job}}

\begin{proposition}\label{prop:experiment_game}
The set of $\chi$-CSE in the job market signaling game of \cite{kubler2008job} that survives the $\chi$-cursed intuitive criterion is characterized as follows:
\begin{itemize}
\item[1.] The separating $\chi$-CSE, in which type $\theta_H$ invests in education while type $\theta_L$ does not, survives the $\chi$-cursed intuitive criterion for $\chi \leq \frac{31}{40}$.
\item[2.] The pooling $\chi$-CSE, in which neither type $\theta_H$ nor type $\theta_L$ invests in education, survives the $\chi$-cursed intuitive criterion for $\chi \geq \frac{11}{20}$.
\item[3.] The hybrid $\chi$-CSE, in which type $\theta_H$ invests in education with probability $\frac{40\chi-22}{9}$ while type $\theta_L$ does not, survives the $\chi$-cursed intuitive criterion for $\frac{11}{20} < \chi < \frac{31}{40}$.
\end{itemize}
\end{proposition}

\noindent\emph{Proof:} Since $\theta_H = 50$ and $\theta_L = 10$, sequential rationality implies 
that the worker's wage will fall within the range $[10, 50]$, regardless of whether 
the worker invests in education.
Therefore, in equilibrium, the type $\theta_L$ worker will never invest in education, 
as the payoff for not investing is at least 10, while the payoff for investing is at most $50 - 45 = 5$.
Building on this observation, we now examine the existence of each type of equilibrium.

\bigskip

\noindent\textbf{Step 1: Separating $\chi$-CSE.} \;By Proposition \ref{prop:curse_bayes_rule}, 
we know that upon observing education, firms' belief that the worker is of type $\theta_H$ is given by
$\mu^\chi(\theta_H|s=1) = 1 - 0.5\chi$. Therefore, conditional on seeing education, firms will offer
a wage of $50(1 - 0.5\chi) + 10(0.5\chi) = 50 - 20\chi$. In contrast, if the worker does not
invest in education, firms' belief is $\mu^\chi(\theta_H|s=0) = 0.5\chi$ and they will offer
a wage of $50(0.5\chi) + 10(1-0.5\chi) = 10 + 20\chi$. Consequently, the IC condition for 
type $\theta_H$ becomes 
$(50 - 20\chi) - 9 \geq 10 + 20\chi \iff \chi \leq 31/40. $

\bigskip

\noindent\textbf{Step 2: Pooling $\chi$-CSE.} \;In the pooling equilibrium, where both types 
choose not to invest in education, firms will offer a wage of 30 upon observing 
no education (on-path). For the off-path event where 
firms observe education, since type $\theta_L$ cannot be better off by choosing
education, the $\chi$-cursed intuitive criterion requires that $T^\chi(s=1) = \{\theta_L \}$ and 
$\mu^\chi(\theta_L | s=1) = 0.5\chi$. In other words, upon observing education, 
firms will offer a wage of $50 - 20\chi$. The IC condition for type $\theta_H$ then becomes
$30 \geq (50 - 20\chi) - 9 \iff \chi \geq 11/20.$

\bigskip
\noindent\textbf{Step 3: Hybrid $\chi$-CSE.} \;Because type $\theta_L$ never invests in 
education in equilibrium, it suffices to consider the hybrid equilibrium in which 
type $\theta_H$ randomizes between investing and not investing. Suppose the type $\theta_H$ worker 
chooses not to invest in education with probability $q$. By Proposition \ref{prop:curse_bayes_rule}, 
firms' beliefs upon observing education and no education are given by
\begin{align*}
    \mu^\chi(\theta_H | s= 0) = 0.5\chi + (1-\chi)\left[\frac{q}{1+q} \right] \;\;\;\;\mbox{and}\;\;\;\; 
    \mu^\chi(\theta_H | s= 1) = 1 - 0.5\chi.
\end{align*}
In equilibrium, type $\theta_H$ must be indifferent between investing and not investing. 
Therefore, 
\begin{align*}
    50\left[0.5\chi + (1-\chi)\left(\frac{q}{1+q} \right) \right] + 10 
    \left[0.5\chi + (1-\chi)\left(\frac{1}{1+q} \right) \right] &= 41 - 20\chi \\
    &\iff q = \frac{31-40\chi}{9},
\end{align*}
and $q\in (0,1) \iff \chi\in \left(\frac{11}{20}, \frac{31}{40} \right)$. This suggests that a hybrid $\chi$-CSE
exists for $\chi\in \left(\frac{11}{20}, \frac{31}{40} \right)$ where type $\theta_H$ invests with probability 
$1-q = \frac{40\chi-22}{9}$. This completes the proof. $\blacksquare$

\subsection*{A Trinary-Choice Extension of \cite{kubler2008job}}

To illustrate how the range of $\chi$ values supporting separating $\chi$-CSE changes with the number of available education levels, we extend the binary-choice setting of \cite{kubler2008job} to a trinary-choice setting with three available education levels, $s \in \left\{0,\frac{2}{3},1\right\}$. Following \cite{kubler2008job}, the productivities of types $\theta_H$ and $\theta_L$ are 50 and 10, respectively, and the cost function is given by $c(s|\theta) = \frac{450s}{\theta}$. Accordingly, the education costs for the two types are summarized in the table below. For simplicity, we focus on pure-strategy $\chi$-CSE in this trinary-choice setting, which are characterized in Proposition \ref{prop:three-choice}.

\begin{table}[H]
\centering
\renewcommand{\arraystretch}{1.25}
\begin{tabular}{ccccc}
 &  & \multicolumn{3}{c}{Education Choices} \\ \cline{2-5} 
\multicolumn{1}{c|}{} & \multicolumn{1}{c|}{Costs} & $s=0$ & $s=\frac{2}{3}$ & \multicolumn{1}{c|}{$s=1$} \\ \cline{2-5} 
\multicolumn{1}{c|}{Types} & \multicolumn{1}{c|}{$\theta_H$} & 0 & 6 & \multicolumn{1}{c|}{9} \\
\multicolumn{1}{c|}{} & \multicolumn{1}{c|}{$\theta_L$} & 0 & 30 & \multicolumn{1}{c|}{45} \\ \cline{2-5} 
\end{tabular}
\end{table}

\begin{proposition}\label{prop:three-choice}
The set of pure-strategy $\chi$-CSE in the job market signaling game of \cite{kubler2008job} with a trinary education choice is characterized as follows:
\begin{itemize}
\item[1.] The strategy profile, in which type $\theta_H$ chooses $s=1$ while type $\theta_L$ chooses $s=0$, is a separating $\chi$-CSE for $\chi \leq \frac{31}{40}$.
\item[2.] The strategy profile, in which type $\theta_H$ chooses $s=\frac{2}{3}$ while type $\theta_L$ chooses $s=0$, is a separating $\chi$-CSE for $
\frac{1}{4}\leq \chi \leq  \frac{17}{20}$.
\item[3.] The strategy profile, in which both types $\theta_H$ and $\theta_L$ choose $s=0$, is a pooling $\chi$-CSE for all $\chi \in [0,1]$.
\end{itemize}

\end{proposition}

\noindent\emph{Proof:} Similar to the proof of Proposition \ref{prop:experiment_game}, 
since $\theta_H = 50$ and $\theta_L = 10$, sequential rationality implies 
that the worker's wage will fall within the range $[10, 50]$, regardless of whether 
the worker invests in education.
Building on this observation, we now examine the existence of each type of equilibrium.

\bigskip

\noindent\textbf{Step 1: Separating $\chi$-CSE ($\theta_H$ 
choosing $s=1$).} \;By the same argument as in Step 1 of the proof of Proposition \ref{prop:experiment_game}, the strategy profile in which type $\theta_H$ chooses $s=1$ while type $\theta_L$ chooses $s=0$ is a separating $\chi$-CSE for $\chi \leq \frac{31}{40}$.

\bigskip

\noindent\textbf{Step 2: Separating $\chi$-CSE ($\theta_H$ 
choosing $s=\frac{2}{3}$).} \;By Proposition \ref{prop:curse_bayes_rule}, 
we know that upon observing $s=\frac{2}{3}$, firms' belief is given by
$\mu^\chi(\theta_H|s=\frac{2}{3}) = 1 - 0.5\chi$. Therefore, conditional on seeing $s=\frac{2}{3}$, firms will offer
a wage of $50 - 20\chi$. In contrast, if the worker chooses $s=0$, firms' belief is $\mu^\chi(\theta_H|s=0) = 0.5\chi$ and they will offer
a wage of $ 10 + 20\chi$. Thus, the IC condition for 
type $\theta_H$ becomes 
$(50 - 20\chi) - 6 \geq 10 + 20\chi \iff \chi \leq 17/20 $, and 
the IC condition for 
type $\theta_L$ is  $10 + 20\chi \geq (50 - 20\chi)-30 \iff \chi \geq 1/4 $. As a result, 
we conclude that this strategy profile is a separating $\chi$-CSE for 
$\frac{1}{4}\leq \chi \leq \frac{17}{20}$.

\bigskip

\noindent\textbf{Step 3: Pooling $\chi$-CSE.} \;In the pooling equilibrium, in which both types choose not to invest in education, firms offer a wage of 30 upon observing no education (on-path). This strategy profile can be supported as a pooling $\chi$-CSE by setting 
firms’ off-path belief to $\mu^\chi(\theta_H | s) = 0.5\chi$. Consequently, neither type has an incentive to deviate, since the off-path wage is $10 + 20\chi \leq 30$ for all $\chi \in [0,1]$.

\bigskip

There is no other pure-strategy $\chi$-CSE because (1) in any separating equilibrium, type $\theta_L$ must choose $s=0$, and (2) there is no other pooling equilibrium, since type $\theta_L$’s cost of choosing any positive education level exceeds the average productivity. This completes the proof. $\blacksquare$

\bigskip

Comparing the trinary-choice extension with the original binary-choice setting, we find that adding the additional education level $s=\frac{2}{3}$ to the message space has no effect on the prediction of the standard intuitive criterion. In both settings, the unique equilibrium that survives the intuitive criterion is the separating equilibrium in which type $\theta_H$ chooses $s=1$ and type $\theta_L$ chooses $s=0$.

In contrast, the predictions of $\chi$-CSE are sensitive to the addition of this extra education level. Specifically, expanding the message space enlarges the range of $\chi$ values for which a separating $\chi$-CSE exists and information transmission can be sustained. In the original binary-choice setting, a separating or hybrid $\chi$-CSE exists only for $\chi \leq \frac{31}{40}$. When the additional education level $s=\frac{2}{3}$ becomes available, Proposition \ref{prop:three-choice} shows that a separating $\chi$-CSE exists for $\chi \leq \frac{17}{20}$. In particular, for $\frac{31}{40} < \chi \leq \frac{17}{20}$, there exists a separating $\chi$-CSE in which type $\theta_H$ uses $s=\frac{2}{3}$, 
a less costly signal, to distinguish itself from type $\theta_L$. This option is unavailable in the binary-choice setting.

\section{Summary of the \cite{kubler2008job} Data}

\begin{table}[htbp!]
\centering
\caption{Average Investment Rates in the SIG2 Treatment}
\label{tab:sig2_investment}
\renewcommand{\arraystretch}{1.15}
\begin{threeparttable}
\begin{tabular}{cccccccccccc}
\hline
 &  &  & \multicolumn{4}{c}{Investment \% of High Type} &  & \multicolumn{4}{c}{Investment \% of Low Type} \\ \cline{4-7} \cline{9-12} 
Block & Periods &  & N & Mean & SD & p-value &  & N & Mean & SD & p-value \\ \hline
1 & 1 -- 8 &  & 37 & 0.378 & 0.492 & $<0.001$ &  & 35 & 0.143 & 0.355 & 0.023 \\
2 & \phantom{1}9 -- 16 &  & 41 & 0.634 & 0.488 & $<0.001$ &  & 31 & 0.290 & 0.461 & 0.002 \\
3 & 17 -- 24 &  & 40 & 0.775 & 0.423 & 0.002 &  & 32 & 0.063 & 0.246 & 0.161 \\
4 & 25 -- 32 &  & 34 & 0.382 & 0.493 & $<0.001$ &  & 38 & 0.026 & 0.162 & 0.324 \\
5 & 33 -- 40 &  & 37 & 0.757 & 0.435 & 0.002 &  & 35 & 0.114 & 0.323 & 0.044 \\
6 & 41 -- 48 &  & 34 & 0.676 & 0.475 & $<0.001$ &  & 38 & 0.000 & 0.000 & --- \\ \hline
All & \phantom{1}1 -- 48 &  & 223 & 0.605 & 0.490 & $<0.001$ &  & 209 & 0.100 & 0.301 & $<0.001$ \\ \hline
\end{tabular}
\begin{tablenotes}
\item The p-values are obtained from two-tailed t-tests comparing the observed
data to the prediction of the intuitive criterion, which predicts that high type workers 
should invest, while low type workers should not invest.
\end{tablenotes}
\end{threeparttable}
\end{table}

\begin{table}[htbp!]
\centering
\caption{Average Investment Rates in the SIG3 Treatment}
\label{tab:sig3_investment}
\renewcommand{\arraystretch}{1.15}
\begin{threeparttable}
\begin{tabular}{cccccccccccc}
\hline
 &  &  & \multicolumn{4}{c}{Investment \% of High Type} &  & \multicolumn{4}{c}{Investment \% of Low Type} \\ \cline{4-7} \cline{9-12} 
Block & Periods &  & N & Mean & SD & p-value &  & N & Mean & SD & p-value \\ \hline
1 & 1 -- 6 &  & 27 & 0.667 & 0.480 & 0.001 &  & 27 & 0.111 & 0.320 & 0.083 \\
2 & \phantom{1}7 -- 12 &  & 32 & 0.563 & 0.504 & $<0.001$ &  & 22 & 0.091 & 0.294 & 0.162 \\
3 & 13 -- 18 &  & 22 & 0.727 & 0.456 & 0.011 &  & 32 & 0.188 & 0.397 & 0.012 \\
4 & 19 -- 24 &  & 30 & 0.800 & 0.407 & 0.012 &  & 24 & 0.208 & 0.415 & 0.022 \\
5 & 25 -- 30 &  & 30 & 0.767 & 0.430 & 0.006 &  & 24 & 0.083 & 0.282 & 0.162 \\
6 & 31 -- 36 &  & 25 & 0.760 & 0.436 & 0.011 &  & 29 & 0.034 & 0.186 & 0.326 \\
7 & 37 -- 42 &  & 28 & 0.571 & 0.504 & $<0.001$ &  & 26 & 0.038 & 0.196 & 0.327 \\
8 & 43 -- 48 &  & 30 & 0.833 & 0.379 & 0.023 &  & 24 & 0.083 & 0.282 & 0.162 \\ \hline
All & \phantom{1}1 -- 48 &  & 224 & 0.710 & 0.455 & $<0.001$ &  & 208 & 0.106 & 0.308 & $<0.001$ \\ \hline
\end{tabular}
\begin{tablenotes}
\item The p-values are obtained from two-tailed t-tests comparing the observed
data to the prediction of the intuitive criterion, which predicts that high type workers 
should invest, while low type workers should not invest.
\end{tablenotes}
\end{threeparttable}
\end{table}

\end{document}